\documentclass[twocolumn, amsmath,amssymb,superscriptaddress,prb]{revtex4}
\usepackage{graphicx}% Include figure files
\usepackage{bm}% bold math
\usepackage{subfigure}
\usepackage{multirow}
\usepackage{color}
\begin{document}

%% Notice placement of commas and superscripts and use of &
%% in the author list

\title{Giant plateau in the {THz} {Faraday} angle in gated {Bi$_2$Se$_3$}}

%Broad vast quantized plateau in the THz Faraday angle in gated

%Observation of a Dirac cone shift in a buried topological interface state
%
%Gated terahertz measurements in a single passivated topological surface state in Bi$_2$Se$_3$ thin films
%THz transport in a single passivated topological surface state in Bi2Se3 films

%transport mass and lifetime measurements in a single topological
%surface state in Bi$_2$Se$_3$ thin films}

%% Gated cyclotron resonance of a single topological surface state in Bi$_2$Se$_3$ thin films}

%% Notice placement of commas and superscripts and use of &
%% in the author list
\author{Gregory S. Jenkins}
    \homepage{http://www.irhall.umd.edu}
    \email{GregJenkins@MyFastMail.com}
    \affiliation{Department of Physics, University of Maryland at College park, College Park, Maryland, 20742, USA}
    \affiliation{Center for Nanophysics and Advanced Materials, University of Maryland at College park, College Park, Maryland, 20742, USA}
\author{Andrei B. Sushkov}
    \affiliation{Department of Physics, University of Maryland at College park, College Park, Maryland, 20742, USA}
    \affiliation{Center for Nanophysics and Advanced Materials, University of Maryland at College park, College Park, Maryland, 20742, USA}
    \affiliation{Materials Research Science and Engineering Center, University of Maryland at College park, College Park, Maryland, 20742, USA}
\author{Don C. Schmadel}
    \affiliation{Department of Physics, University of Maryland at College park, College Park, Maryland, 20742, USA}
    \affiliation{Center for Nanophysics and Advanced Materials, University of Maryland at College park, College Park, Maryland, 20742, USA}
\author{M.-H. Kim}
    \affiliation{Department of Physics, University of Maryland at College park, College Park, Maryland, 20742, USA}
    \affiliation{Center for Nanophysics and Advanced Materials, University of Maryland at College park, College Park, Maryland, 20742, USA}
\author{Matthew Brahlek}
    \affiliation{Department of Physics and Astronomy, The State University of New Jersey, Piscataway, New Jersey 08854, USA}
\author{Namrata Bansal}
    \affiliation{Department of Physics and Astronomy, The State University of New Jersey, Piscataway, New Jersey 08854, USA}
\author{Seongshik Oh}
    \affiliation{Department of Physics and Astronomy, The State University of New Jersey, Piscataway, New Jersey 08854, USA}
\author{H. Dennis Drew}
    \affiliation{Department of Physics, University of Maryland at College park, College Park, Maryland, 20742, USA}
    \affiliation{Center for Nanophysics and Advanced Materials, University of Maryland at College park, College Park, Maryland, 20742, USA}
    \affiliation{Materials Research Science and Engineering Center, University of Maryland at College park, College Park, Maryland, 20742, USA}
\date{\today}

\begin{abstract}
We report gated terahertz Faraday angle measurements on epitaxial Bi$_2$Se$_3$ thin films capped with In$_2$Se$_3$. A plateau is observed in the real part of the Faraday angle at an onset gate voltage corresponding to no band bending at the surface which persists into accumulation. The plateau is two orders of magnitude flatter than the step size expected from a single Landau Level in the low frequency limit,  quantized in units of the fine structure constant. At $8$ T, the plateau extends over a range of gate voltage that spans an electron density greater than 14 times the quantum flux density. Both the imaginary part of the Faraday angle and transmission measurements indicate dissipative off-axis and longitudinal conductivity channels associated with the plateau.
\end{abstract}

% }
\maketitle

In two-dimensional quantum Hall systems, one-dimensional edge states are topologically protected quantum states characterized by integer Chern numbers. Three dimensional topological insulators are characterized by a quantized magneto-electric term appearing in the effective Lagrangian, quantized in units of the fine structure constant $\alpha=\frac{e^2}{\hbar c}$.\cite{1985QHE-FA-alpha,HasanKaneRMP2010,QiZhangRMP2011,Maciejko2010,TseMacDonald2010} The quantization condition of this magneto-electric term stems from a Chern-Simons form of the Berry phase, known as the axion angle in particle physics.\cite{wilczek1987, QiZhang2008}  In the non-trivial case, the axion angle attains odd-multiple values of $\pi$ resulting in two dimensional topologically protected surface states that are spin polarized. By breaking time reversal symmetry, these singly-occupied two dimensional Dirac states are predicted to exhibit a $1/2$-quantized Hall step near the Dirac point, a smoking gun signature of the topological origin of the surface state.\cite{QiZhang2008} Kerr or Faraday terahertz (THz) measurements offer a method to cleanly measure this quantized step since no contacts to the sample are required alleviating complications that can occur from gating a sample with no edges.\cite{Jenkins2012,Jenkins_PRB2010}

Bi$_2$Se$_3$ has a single surface-state Dirac cone whose Dirac point is well above the valence band. The material has one of the largest bulk band gaps of all known topological insulators allowing access to the Dirac cone over a large range of Fermi energies. However, Bi$_2$Se$_3$ is susceptible to n-type bulk doping from defects as well as environmental surface doping.\cite{ButchPRB2010, Jenkins_PRB2010, KongARPESNN2011, Checkelsky_OngPRL2011, Analytis_NP2010, SteinbergHerreroNNano2010, Dohun_Fuhrer2012} High quality epitaxially grown thin films of Bi$_2$Se$_3$ have relatively low n-type bulk doping,\cite{Oh_arxiv2011} and subsequent capping with In$_2$Se$_3$ further lowers the Fermi level of the surface states allowing conventional gating to reach the Dirac point.\cite{Jenkins2012}

\begin{figure}
\includegraphics[scale=.4]{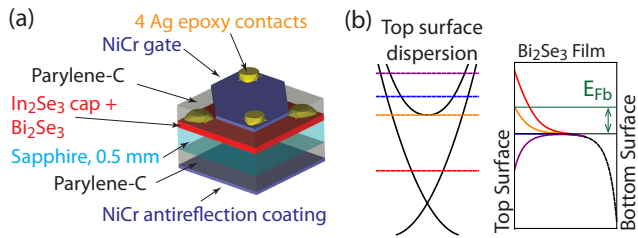}% Here is how to import EPS art
\caption{\label{fig1}\textbf{Sample layout and band bending diagram:} (a) A Bi$_2$Se$_3$ 60 (40) QL film is capped with $10$ nm of In$_2$Se$_3$, and encased in  $590$ nm (620 nm) of Parylene-C. A NiCr gate $\sim400$  $\Omega/\square$ and antireflection coating $\sim275$ $\Omega/\square$ were deposited by evaporation. Pairs of silver epoxy contacts were made to the Bi$_2$Se$_3$ film and NiCr gate. (b) The band bending diagram on the right side depicts the conduction band edge from large negative gate voltages (red) to large positive gate voltages (purple) throughout the cross section of the film. The left diagram depicts the top topological surface state and associated bulk band (shown as a parabola) dispersion, and the dashed colored horizontal lines represent the associated surface Fermi levels due to the band bending. The specific Fermi levels at the surface depicted in blue and orange are referenced in the text as the flat band and conduction band edge (CBE) conditions, respectively.}
\end{figure}

No previous optical probe has been reported that attempts to measure the quantized Hall effect on a topological insulator as a function of gate voltage.\cite{Jenkins2012, Jenkins_PRB2010, Sushkov_PRB2010, Molenkamp2011, HancockMolenkamp2011, KvonGanichev2012, SchafgansBasov2012, AguilarPRL2012} We report Faraday angle and transmission measurements performed at a fixed laser frequency of $\omega/2\pi=0.74$ THz as a function of gate voltage at discrete magnetic fields up to $8$ T on two topological insulating films. These Bi$_2$Se$_3$ thin films, $40$ and $60$ quintuple layers thick (1 QL $\approx$ 1 nm), were grown epitaxially onto $0.5$ mm thick sapphire substrates and capped with $10$ nm thick In$_2$Se$_3$ layers without breaking vacuum.\cite{Oh_arxiv2011,Bansal_OhThinFilm2011} As depicted in Figure \ref{fig1}(a), the dielectric Parylene-C was deposited encasing the sample to a thickness of $590$ nm ($620$ nm) for the 60 (40) QL film. A NiCr top gate and a NiCr antireflection coating were evaporated onto the Parylene-C. For the 60 QL device, the gate moves a charge density of $2.6 \times 10^{10} e/\text{cm}^2$ per volt on or off the  Bi$_2$Se$_3$ film.

The Faraday angle is related to the off-axis conductivity $\sigma_{xy}$  of a thin film  by $\theta_F \approx Z \sigma_{xy} /(1+Z \sigma_{xx})$ where $Z=Z_0/(n_s+1)$, and $n_s$ is the substrate index of refraction, $Z_0$ is the impedance of free space,  and $\sigma_{xx}$ is the longitudinal conductivity.\cite{Jenkins_RSI_2010, Jenkins2012} Considering the case where the cyclotron frequency $\omega_c$ is large compared to the scattering rate $\gamma$ and radiation frequency $\omega$, and $Z \sigma_{xx}$ is small compared with $1$, crossing a Landau level results in $\Delta \theta_F\approx 2 \alpha / (n_s + 1)$. Under very specific geometric conditions, the substrate properties can be ignored and $\Delta \theta_F$ is quantized exactly in units of $\alpha$.\cite{OConnellWallace1982, 1985QHE-FA-alpha, Maciejko2010, TseMacDonald2010} More generally, the quantization of $\Delta \theta_F$ from step-to-step is scaled by the material properties of the substrate and any longitudinal conductivity contributions (like a conducting gate or other dissipation channels), as well as corrections from non-zero $\gamma$ and $\omega$.\cite{MorimotoAoki2009, StierCerne2011}

The complex Faraday angle measurement at THz frequencies utilizes polarization-modulated fixed-frequency laser radiation, a technique detailed elsewhere.\cite{Jenkins_RSI_2010} The beam is normally incident on the film held at $10$ K that is maintained in a normally applied magnetic field. The  transmission through the film is measured concurrently. The differential Faraday angle $\Delta \theta_F$ and the differential transmission $\Delta T$ are differences between the optical signals measured at a fixed gate voltage $V_g^0=+170$ V and the signals measured at a variable gate voltage $V_g$.

\begin{figure*}
\includegraphics[scale=.6]{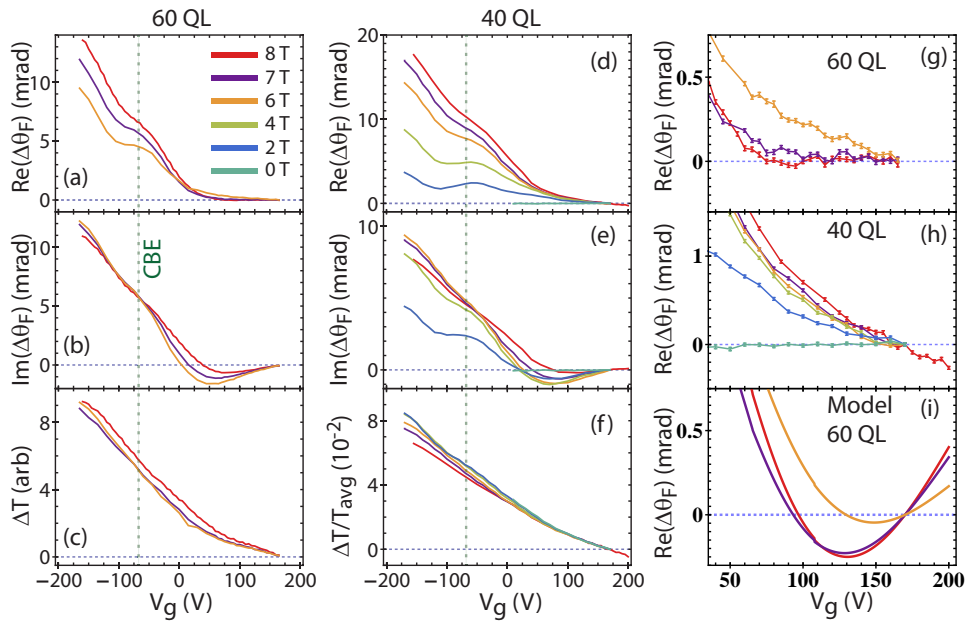}% Here is how to import EPS art
\caption{\label{fig2}\textbf{$\Delta \theta_F$ and $\Delta T$ data and model:} All transmission and Faraday angle measurements were performed at $\omega/2\pi = 0.74$ THz at discrete magnetic fields that are color coded according to the key in panel (a). Differential optical measurements are differences between two different gate voltages, one is fixed at $V_g^0=170$ V and the other $V_g$ is variable. Maximum reported gate voltages are near the break down limit of the devices. The differential transmission $\Delta T$ is normalized to the average transmission $T_{avg}$ of the two gate values only in panel (f). The conduction band edge (CBE) condition at the top surface is demarcated with a vertical dotted line in all figures. (a-c) Differential measurements of the Faraday angle and transmission are shown for the 60 QL film maintained at $10$ K and (d-f) the 40 QL film maintained at $10$ K. (g,h) A zoomed view of the Re$(\Delta \theta_F$) for the 60 and 40 QL films, respectively. (i) A model of the Re$(\Delta \theta_F$) for the 60 QL device incorporating the Drude parameters that reproduce all the optical data in reference \citenum{Jenkins2012}.}
\end{figure*}

The differential Faraday angle $\Delta \theta_F$ and differential transmission $\Delta T$ as a function of gate voltage at fixed fields for the two devices are reported in Figure \ref{fig2}. Measurements performed at THz frequencies on the same 60 quintuple layer (QL) device, but as a function of magnetic field, are reported in reference \citenum{Jenkins2012}. The gate sweeps at fixed magnetic field presented here offer better signal-to-noise than previous measurements. $Re(\Delta \theta_F)$ is measured to within a standard deviation of the mean of $30$ $\mu$rad without systematic drift effects associated with changing the magnetic field. With this additional sensitivity, $Re(\Delta \theta_F)$ of the $60$ QL film shows an extraordinarily flat response over a tremendous span of gate voltages, from $+170$ V to $+70$ V.

The previous THz measurements on the 60 QL film provide an overview of the characteristics of the film. The salient features of the THz data are captured utilizing two gate dependent Drude terms in the conductivity: the first term is primarily associated with the bulk carriers of the gate-modulated screened region near the top surface (MTB), and the second term is the top topological surface state (TSS). The conception of the model is based upon the Thomas-Fermi band bending in the film, depicted in Figure \ref{fig1}(b). All carrier contributions in the film were measured to be n-type. Based on the small measured cyclotron mass $m_c=\hbar k_F/v_F$ at high negative gate voltages, the Dirac point is estimated to be in the vicinity of $-170$ V.  The conduction band edge (CBE) voltage location is $V_g=-70$ V deduced from a sudden increase in the measured surface state scattering rate, interpreted as the optical signature where surface state carriers begin scattering into bulk channels, and is demarcated by the green vertical dashed lines in Figure \ref{fig2}.\cite{Jenkins2012} This corresponds to a CBE that is 80 meV above the Dirac point, whereas the vacuum/Bi$_2$Se$_3$ interface CBE characterized by surface probes is at 190 meV\cite{XiaARPESNP2009, ZhuPRL2011, Bahramy_Hofmann_2012} implying a shift in the Dirac point due to the In$_2$Se$_3$/Bi$_2$Se$_3$ interface properties.\cite{Jenkins2012} With regard to the plateau region, this is a minor point since top surface bulk carriers are degenerate with the surface state carriers over the full range of the plateau whether the CBE location is 80 or 190 meV above the Dirac point (as discussed in more detail in the Appendix).

There are a number of important observations associated with the plateau in the $Re(\Delta \theta_F)$ shown in Figure \ref{fig2}(a) and (g). The first is the degree of flatness.  Such an extremely flat response is very difficult to reproduce invoking non-quantized behavior. In the simplest view, a gate moves electrons to the film with increasingly positive gate voltage, predictably lowering the transmission as reported in Figure \ref{fig2}(c), yet the real part of the Faraday angle remains constant over a large range of voltages.

A slightly more sophisticated approach using the same multi-fluid Drude model that reproduced the features of all the data reported in reference \citenum{Jenkins2012} does not reproduce the extremely flat behavior of the $Re(\Delta \theta_F)$ gate sweeps in fixed magnetic field, as shown in Figure \ref{fig2}(i). The gate dependent scattering rates of the TSS and MTB Drude terms give rise to an extremum in the modelled response of Figure \ref{fig2}(i). The observation in Figure \ref{fig2}(g) of such a flat response over such a large range of gate voltages is irreconcilable with
two gate-dependent n-type Drude terms while maintaining consistency with previous optical data. The data suggests the existence of localized states associated with $\sigma_{xy}$ which absorb the transferred charge from the gate but give no additional contribution to $Re(\theta_F)$.

The formation of Landau levels is one way that can provide localized states. Although we currently do not have a complete explanation for the observed plateau, recent dc Hall measurements on highly doped $\gtrsim 10^{19}$ cm$^{-3}$ Bi$_2$Se$_3$ bulk crystals, performed on many different samples spanning four orders of magnitude in thickness, appear to support the scenario that an extraordinary bulk quantum Hall effect (QHE) occurs in which each pair of QLs contribute a distinct Landau Level to the Hall response.\cite{CaoChen2012} These Landau levels appear to act in unison such that the step-to-step height $\Delta \sigma_{xy} = \text{(\# of QLs)} \frac{e^2}{h}$.  However, $\rho_{xx}$ oscillates but does not go to zero, remaining comparable with $\rho_{xy}$ even within plateaus, indicating appreciable dissipation. Within this context, we discuss the character of the observed plateau in the $Re(\Delta \theta_F)$ shown in Figure \ref{fig2}(g).

For small longitudinal conductivity ($Z \, \sigma_{xx}<<1)$ and large cyclotron frequency $\omega_c>>\gamma,\, \omega$, the finite-frequency off-axis conductivity $\sigma_{xy} = \Omega_0 \omega_c / ((\gamma - \imath \omega)^2 + \omega_c^2)$ reduces to $\sigma_{xy}= n e/B$ since the spectral weight $\Omega_0 = n e^2/m$ and the cyclotron frequency $\omega_c= e B /m_c$. The electron density $n$ changes by $B/(h/e)$ when the Fermi level sweeps through a single Landau Level at fixed magnetic field $B$, so $Z_0 \Delta \sigma_{xy} = 2 \alpha$.  Since sapphire has an index of refraction of $n\approx 3$, $Z_0/R \approx 1$ for the NiCr gate, and $Z_0 \sigma_{xx} \approx 1$, the change in the real part of the Faraday angle upon stepping through a Landau Level is expected to be $Re(\Delta \theta_F) \approx \alpha/3 = 2.4$ mrad in the low frequency limit. In this limit, the $Re(\Delta \theta_F)$ at $8$ T is flat over a range of at least $100$ V to within 1/80 of the expected step size from a single Landau level. The expected step size for a 2DEG has been predicted to increase from this estimate due to the finite probe frequency ($\omega=4.6$ THz) and scattering rate ($1/\tau \approx 5$ THz) in comparison to the cyclotron frequency ($\omega_c=6.4$ THz at $8$ T).\cite{MorimotoAoki2009}

The plateau at positive gate voltage decreases in width with decreasing magnetic field qualitatively consistent with Landau level behavior. The surface state is degenerate with surface accumulated bulk carriers over the full range of the plateau. The step-like feature at $-70$ V at the CBE location is a result of the changing surface state scattering rate.\cite{Jenkins2012} This feature does not display the expected behavoir by a quantized Hall step upon changing the magnetic field.

The width of the positive-voltage plateau is much larger than expected from a single Landau level. The gate moves $2.6 \times 10^{12} $ $e$/cm$^2$ over $100$ V, and each Landau level is occupied by an electron density of $B/(h/e) = 1.9 \times 10^{11}$ cm$^{-2}$ in $8$ T.  At least $14$ Landau levels are traversed over the span of the plateau.

Given that no Landau level steps are discernable in the data at negative gate voltages, and the topological surface state is singly occupied contributing at most 1 Landau level of charge, the conclusion is that the plateau must primarily be a bulk effect.

Figure \ref{fig2}(d-f) and (h) show gated THz measurements performed on a 40 QL Bi$_2$Se$_3$ film. Although the qualitative behavior is very similar to the 60 QL film, there are no plateaus in the $Re(\Delta \theta_F)$ response.  The Thomas-Fermi screening length is estimated to be about $11$ nm based upon dc measurements of bulk densities of similarly prepared films performed over a wide range of thicknesses (see Appendix). Gating the top surface of the $40$ nm film will affect the accumulated bottom surface layer much more than the thicker $60$ nm film. An additional gate dependent conductivity channel associated with the bottom surface accumulation layer can give rise to a different behavior. Study of thicker films is necessary to verify this thickness effect.

The gate dependence of $\Delta T$ and Im($\Delta \theta_F$) of the 60 QL film over the same voltage range where the plateau exists indicates dissipation in the $\sigma_{xx}$ and $\sigma_{xy}$ channels. In this regard, the similarities between the reported dc measured bulk QHE\cite{CaoChen2012} and the gated THz measurements are striking. These dc data show plateaus in the Hall conductivity, consistent with the sum of many Landau level contributions, in the presence of significant dissipative channels.

However, even the highest positive applied gate voltage of $+170$ V, which causes the conduction band to bend downward at the surface as depicted in Figure \ref{fig1}(b), only results in a Fermi level of $\sim40$ meV above the conduction band edge at the surface (see Appendix). This Fermi level translates into a bulk density of $\sim 3 \times 10^{18}$ cm$^{-3}$, about an order of magnitude less than the dc measurements on bulk crystals reporting bulk QHE.\cite{CaoChen2012}

More important, the most significant difference between the THz and dc experiments is associated with the gradient of the potential from band bending effects in the film produced by the applied gate. In the case of the dc measurements, identical independent two-dimensional layers exhibiting the QHE are known to summate in the expected way, corresponding to a single Landau level but with a degeneracy corresponding to the number of layers.\cite{Haavasoja1984} However, if the potential varies appreciably from layer to layer, the Landau levels would not remain degenerate nor give well defined large steps.  In our case, screening effects are expected to cause such a misalignment of Landau levels. At $+170$ V, the bands are bent downward by $\sim30$ meV over $\sim14$ QL (the screening length), producing $\sim2$ meV change in potential energy from QL to QL. The Landau level spacing (cyclotron frequency) in $8$ T is only $\sim 4$ meV, so one might expect the variation of potential would wipe out any additive effect from independent Landau levels that could produce a wide Hall plateau.

This may indicate that the contributing Landau levels are not independent, but interact in a way that tends to lock the levels together. For the dc measured bulk QHE of reference \citenum{CaoChen2012}, the scenario put forth was that the quintuple layers of Bi$_2$Se$_3$ decouple, forming independent 2-DEGs.  Many questions remain about the nature of this decoupling since Shubnikov-de Haas measurements show a 3-D Fermi surface.\cite{Köhler1973,EtoAndo2010}  It is also puzzling how the QHE survives the large measured $\sigma_{xx}$ in the Hall plateaus.\cite{CaoChen2012}  Similarly in this paper we report the phenomenology of an extremely flat Hall response observed at sub-THz frequencies, manifesting in the real part of the Faraday angle that is flat to within 1/80 of the expected low-frequency quantized step size for a single Landau level. This plateau extends over a large range of gate induced carrier density change of $\sim 3 \times 10^{12} \text{cm}^{-2}$, occurring above the conduction band edge, corresponding to more than $14$ occupied Landau levels. A full understanding of this plateau is likely tied to understanding the bulk QHE, which currently remains elusive.

The authors thank M. S. Fuhrer, T. D. Stanescu, and S. Das Sarma for helpful conversations. The work at the University of Maryland is supported by NSF (DMR-1104343) and CNAM. The Rutgers work is supported by IAMDN of Rutgers University, National Science Foundation (NSF DMR-0845464,) and Office of Naval Research (ONR N000140910749).

\section*{Appendix}

The following is a summary of characterization results of the 60 nm film stated in the main text. This particular 60 nm film has been extensively characterized spectroscopically in reference \citenum{Jenkins2012}. Results directly obtained from raw optical data in reference \citenum{Jenkins2012} and dc transport measurements on similarly grown films in reference \citenum{Oh_arxiv2011}, used in conjunction with Thomas-Fermi screening, provide important information of the carriers involved in the gated Faraday angle response in the plateau region.

DC transport measurements on uncapped films, grown in exactly the same manner as our film but without a capping layer, show a carrier density $\thickapprox 4 \times 10^{13}$ cm$^{-2}$ over a wide range of film thicknesses. The thickness independence of the carrier density indicates the bulk contribution is small compared to the two accumulated surfaces (topological surface states and bulk accumulation layers). The bulk carrier density is measured to be $\lesssim 5 \times 10^{17}$ cm$^{-3}$.\cite{Oh_arxiv2011} The cyclotron mass is $\sim0.2$ m$_0$ (taken directly from the THz Faraday and cyclotron resonance data).\cite{Jenkins2012} The screening length $\lambda_s$ depends weakly on bulk density, so it is $\sim 11$ nm. The bulk density corresponds to a bulk Fermi level of $\sim11$ meV.

For comparison, this bulk carrier density distributed uniformly over a $60$ nm film results in a 2-D carrier density of only $\lesssim 3 \times 10^{12}$ cm$^{-2}$. A single-fluid fit to zero-gate FTIR data and Faraday angle give the total carrier density in the film, $\approx1.2\times10^{13}$ cm$^{-2}$.\cite{Jenkins2012}

A very small cyclotron mass measured in the gate-modulated THz cyclotron resonance ($\Delta$-CR) data at high negative gate voltage $V_g$ is obtained directly from the raw data in Figure 2(a) of reference \citenum{Jenkins2012}. Such a small mass is direct evidence of the topological surface state. An immediate consequence is that the Fermi level in the surface state is very near the Dirac point at $V_g=-170$ V. At this voltage, a depletion region necessarily exists at the top surface as indicated by the red curve in Figure \ref{fig1}(b) of the main text.

A bulk density of $n_0=5\times 10^{17}$ cm$^{-3}$ is consistent with dc characterizations\cite{Oh_arxiv2011} and optical characterizations.\cite{Jenkins2012} At $V_g =-170$ V, the Dirac cone is empty and there is a depletion layer of charge $e n_d$ in the film. The gate is capable of moving $n_g = (2.6\times10^{10} \text{cm}^{-2}) \Delta V_g$ carriers for a change in gate voltage $\Delta V_g$. The conduction band is reached when the gate supplies enough charge to fill both the surface state ($e n_{ss}$) up to the conduction band edge as well as the depletion charge, or $n_g =n_{ss} + n_d$. This condition is depicted by the orange curve in Figure \ref{fig1}.

The depletion charge is calculated by solving the Thomas-Fermi screening model, giving $n_d = n_0(\lambda_s^2 + \frac{2 \epsilon}{n_0 e^2} E_c)^{1/2} - n_0 \lambda_s$, where $E_c$ is the energy of the conduction band edge above the Dirac point. From the ARPES measured surface state, $n_{ss}$ is given by the average dispersion $E
= 1.9 k + 12.6 k^2+ 2300 k^6$.\cite{Jenkins2012}

For a conduction band edge that is 190 meV above the Dirac point, $n_{ss} =3.8\times10^{12}$ cm$^{-2}$ and $n_d=2.7\times10^{12}$ cm$^{-2}$. The surface Fermi level will equal the conduction band edge, therefore, when $V_g=+70$ V. If the gate changes another 100 V to $V_g=+170$ V, the top surface becomes accumulated, as depicted in purple in Figure \ref{fig1}.

Optical characterizations of the 60 nm film show evidence that the conduction band edge is at -70 V, as mentioned in the main text. Since the Dirac point is at -170 V, the number of carriers transferred by the gate is $n_g=2.6\times10^{12}$ cm$^{-2}$. Solving $n_g=n_{ss}(E_c) + n_d(E_c)$ gives $E_c=80$ meV where $n_{ss}=1\times10^{12}$ cm$^{-2}$ and $n_d=1.6\times10^{12}$ cm$^{-2}$. To then change the gate to $V_g=+170$ V will transfer an additional $6.2\times10^{12}$ cm$^{-2}$ carriers to the film. Over the entire voltage range in which the plateau is gated, the top surface bulk is accumulated. The resulting surface Fermi level at $V_g=+170$ V is $\approx40$ meV above the conduction band edge, found from solving the Thomas-Fermi screening model (a method that is derived in detail in reference \citenum{Jenkins2012} in the Supplemental Materials, Section II).

In either case, whether the conduction band edge is located at 190 meV or 80 meV above the Dirac point, there are bulk carriers that are degenerate with surface state carriers over the full range of the observed plateau.

Also, a bottom surface accumulation layer exists in both these cases. From $V_g=-170$ V to $V_g=0$,  $4.4\times10^{12}$ cm$^{-2}$ carriers are added to the film filling the depletion layer and transferring carriers into the surface state and bulk.  Therefore, the net carrier density at the top surface is $4.4\times10^{12} \text{cm}^{-2} - n_d$. In both cases, the net carrier density at the top surface is $<3\times 10^{12}$ cm$^{-2}$. The remaining  $> 9\times10^{12}$ cm$^{-2}$ carriers optically characterized by zero-gate measurements\cite{Jenkins2012} are in the bottom surface and bulk. The bulk is a small contribution, implying that the bottom surface is accumulated.

\bibliography{BiSefilmBib}

\begin{thebibliography}{34}
\expandafter\ifx\csname natexlab\endcsname\relax\def\natexlab#1{#1}\fi
\expandafter\ifx\csname bibnamefont\endcsname\relax
  \def\bibnamefont#1{#1}\fi
\expandafter\ifx\csname bibfnamefont\endcsname\relax
  \def\bibfnamefont#1{#1}\fi
\expandafter\ifx\csname citenamefont\endcsname\relax
  \def\citenamefont#1{#1}\fi
\expandafter\ifx\csname url\endcsname\relax
  \def\url#1{\texttt{#1}}\fi
\expandafter\ifx\csname urlprefix\endcsname\relax\def\urlprefix{URL }\fi
\providecommand{\bibinfo}[2]{#2}
\providecommand{\eprint}[2][]{\url{#2}}

\bibitem[{\citenamefont{{Volkov} and {Mikhailov}}(1985)}]{1985QHE-FA-alpha}
\bibinfo{author}{\bibfnamefont{V.~A.} \bibnamefont{{Volkov}}} \bibnamefont{and}
  \bibinfo{author}{\bibfnamefont{S.~A.} \bibnamefont{{Mikhailov}}},
  \bibinfo{journal}{JETP Letters} \textbf{\bibinfo{volume}{41}},
  \bibinfo{pages}{389} (\bibinfo{year}{1985}).

\bibitem[{\citenamefont{Hasan and Kane}(2010)}]{HasanKaneRMP2010}
\bibinfo{author}{\bibfnamefont{M.~Z.} \bibnamefont{Hasan}} \bibnamefont{and}
  \bibinfo{author}{\bibfnamefont{C.~L.} \bibnamefont{Kane}},
  \bibinfo{journal}{Reviews of Modern Physics} \textbf{\bibinfo{volume}{82}},
  \bibinfo{pages}{3045} (\bibinfo{year}{2010}).

\bibitem[{\citenamefont{Qi and Zhang}(2011)}]{QiZhangRMP2011}
\bibinfo{author}{\bibfnamefont{X.-L.} \bibnamefont{Qi}} \bibnamefont{and}
  \bibinfo{author}{\bibfnamefont{S.-C.} \bibnamefont{Zhang}},
  \bibinfo{journal}{Reviews of Modern Physics} \textbf{\bibinfo{volume}{83}},
  \bibinfo{pages}{1057} (\bibinfo{year}{2011}).

\bibitem[{\citenamefont{Maciejko et~al.}(2010)\citenamefont{Maciejko, Qi, Drew,
  and Zhang}}]{Maciejko2010}
\bibinfo{author}{\bibfnamefont{J.}~\bibnamefont{Maciejko}},
  \bibinfo{author}{\bibfnamefont{X.-L.} \bibnamefont{Qi}},
  \bibinfo{author}{\bibfnamefont{H.~D.} \bibnamefont{Drew}}, \bibnamefont{and}
  \bibinfo{author}{\bibfnamefont{S.-C.} \bibnamefont{Zhang}},
  \bibinfo{journal}{Physical Review Letters} \textbf{\bibinfo{volume}{105}},
  \bibinfo{pages}{166803} (\bibinfo{year}{2010}).

\bibitem[{\citenamefont{Tse and MacDonald}(2010)}]{TseMacDonald2010}
\bibinfo{author}{\bibfnamefont{W.-K.} \bibnamefont{Tse}} \bibnamefont{and}
  \bibinfo{author}{\bibfnamefont{A.~H.} \bibnamefont{MacDonald}},
  \bibinfo{journal}{Physical Review Letters} \textbf{\bibinfo{volume}{105}},
  \bibinfo{pages}{057401} (\bibinfo{year}{2010}).

\bibitem[{\citenamefont{Wilczek}(1987)}]{wilczek1987}
\bibinfo{author}{\bibfnamefont{F.}~\bibnamefont{Wilczek}},
  \bibinfo{journal}{Physical Review Letters} \textbf{\bibinfo{volume}{58}},
  \bibinfo{pages}{1799} (\bibinfo{year}{1987}),
  \urlprefix\url{http://link.aps.org/doi/10.1103/PhysRevLett.58.1799}.

\bibitem[{\citenamefont{Qi et~al.}(2008)\citenamefont{Qi, Hughes, and
  Zhang}}]{QiZhang2008}
\bibinfo{author}{\bibfnamefont{X.-L.} \bibnamefont{Qi}},
  \bibinfo{author}{\bibfnamefont{T.~L.} \bibnamefont{Hughes}},
  \bibnamefont{and} \bibinfo{author}{\bibfnamefont{S.-C.} \bibnamefont{Zhang}},
  \bibinfo{journal}{Physical Review B} \textbf{\bibinfo{volume}{78}},
  \bibinfo{pages}{195424} (\bibinfo{year}{2008}).

\bibitem[{\citenamefont{Jenkins et~al.}(2012)\citenamefont{Jenkins, Sushkov,
  Schmadel, Bichler, Koblmueller, Brahlek, Bansal, Oh, and Drew}}]{Jenkins2012}
\bibinfo{author}{\bibfnamefont{G.~S.} \bibnamefont{Jenkins}},
  \bibinfo{author}{\bibfnamefont{A.~B.} \bibnamefont{Sushkov}},
  \bibinfo{author}{\bibfnamefont{D.~C.} \bibnamefont{Schmadel}},
  \bibinfo{author}{\bibfnamefont{M.}~\bibnamefont{Bichler}},
  \bibinfo{author}{\bibfnamefont{G.}~\bibnamefont{Koblmueller}},
  \bibinfo{author}{\bibfnamefont{M.}~\bibnamefont{Brahlek}},
  \bibinfo{author}{\bibfnamefont{N.}~\bibnamefont{Bansal}},
  \bibinfo{author}{\bibfnamefont{S.}~\bibnamefont{Oh}}, \bibnamefont{and}
  \bibinfo{author}{\bibfnamefont{H.~D.} \bibnamefont{Drew}},
  \bibinfo{journal}{arXiv:1208.3881}  (\bibinfo{year}{2012}),
  \urlprefix\url{http://arxiv.org/abs/1208.3881}.

\bibitem[{\citenamefont{Jenkins
  et~al.}(2010{\natexlab{a}})\citenamefont{Jenkins, Sushkov, Schmadel, Butch,
  Syers, Paglione, and Drew}}]{Jenkins_PRB2010}
\bibinfo{author}{\bibfnamefont{G.~S.} \bibnamefont{Jenkins}},
  \bibinfo{author}{\bibfnamefont{A.~B.} \bibnamefont{Sushkov}},
  \bibinfo{author}{\bibfnamefont{D.~C.} \bibnamefont{Schmadel}},
  \bibinfo{author}{\bibfnamefont{N.~P.} \bibnamefont{Butch}},
  \bibinfo{author}{\bibfnamefont{P.}~\bibnamefont{Syers}},
  \bibinfo{author}{\bibfnamefont{J.}~\bibnamefont{Paglione}}, \bibnamefont{and}
  \bibinfo{author}{\bibfnamefont{H.~D.} \bibnamefont{Drew}},
  \bibinfo{journal}{Physical Review B} \textbf{\bibinfo{volume}{82}},
  \bibinfo{pages}{125120} (\bibinfo{year}{2010}{\natexlab{a}}).

\bibitem[{\citenamefont{Butch et~al.}(2010)\citenamefont{Butch, Kirshenbaum,
  Syers, Sushkov, Jenkins, Drew, and Paglione}}]{ButchPRB2010}
\bibinfo{author}{\bibfnamefont{N.~P.} \bibnamefont{Butch}},
  \bibinfo{author}{\bibfnamefont{K.}~\bibnamefont{Kirshenbaum}},
  \bibinfo{author}{\bibfnamefont{P.}~\bibnamefont{Syers}},
  \bibinfo{author}{\bibfnamefont{A.~B.} \bibnamefont{Sushkov}},
  \bibinfo{author}{\bibfnamefont{G.~S.} \bibnamefont{Jenkins}},
  \bibinfo{author}{\bibfnamefont{H.~D.} \bibnamefont{Drew}}, \bibnamefont{and}
  \bibinfo{author}{\bibfnamefont{J.}~\bibnamefont{Paglione}},
  \bibinfo{journal}{Physical Review B} \textbf{\bibinfo{volume}{81}},
  \bibinfo{pages}{241301} (\bibinfo{year}{2010}).

\bibitem[{\citenamefont{Kong et~al.}(2011)\citenamefont{Kong, Cha, Lai, Peng,
  Analytis, Meister, Chen, Zhang, Fisher, Shen et~al.}}]{KongARPESNN2011}
\bibinfo{author}{\bibfnamefont{D.}~\bibnamefont{Kong}},
  \bibinfo{author}{\bibfnamefont{J.~J.} \bibnamefont{Cha}},
  \bibinfo{author}{\bibfnamefont{K.}~\bibnamefont{Lai}},
  \bibinfo{author}{\bibfnamefont{H.}~\bibnamefont{Peng}},
  \bibinfo{author}{\bibfnamefont{J.~G.} \bibnamefont{Analytis}},
  \bibinfo{author}{\bibfnamefont{S.}~\bibnamefont{Meister}},
  \bibinfo{author}{\bibfnamefont{Y.}~\bibnamefont{Chen}},
  \bibinfo{author}{\bibfnamefont{H.-J.} \bibnamefont{Zhang}},
  \bibinfo{author}{\bibfnamefont{I.~R.} \bibnamefont{Fisher}},
  \bibinfo{author}{\bibfnamefont{Z.-X.} \bibnamefont{Shen}},
  \bibnamefont{et~al.}, \bibinfo{journal}{ACS Nano}
  \textbf{\bibinfo{volume}{5}}, \bibinfo{pages}{4698} (\bibinfo{year}{2011}).

\bibitem[{\citenamefont{Checkelsky et~al.}(2011)\citenamefont{Checkelsky, Hor,
  Cava, and Ong}}]{Checkelsky_OngPRL2011}
\bibinfo{author}{\bibfnamefont{J.~G.} \bibnamefont{Checkelsky}},
  \bibinfo{author}{\bibfnamefont{Y.~S.} \bibnamefont{Hor}},
  \bibinfo{author}{\bibfnamefont{R.~J.} \bibnamefont{Cava}}, \bibnamefont{and}
  \bibinfo{author}{\bibfnamefont{N.~P.} \bibnamefont{Ong}},
  \bibinfo{journal}{Physical Review Letters} \textbf{\bibinfo{volume}{106}},
  \bibinfo{pages}{196801} (\bibinfo{year}{2011}).

\bibitem[{\citenamefont{Analytis et~al.}(2010)\citenamefont{Analytis, McDonald,
  Riggs, Chu, Boebinger, and Fisher}}]{Analytis_NP2010}
\bibinfo{author}{\bibfnamefont{J.~G.} \bibnamefont{Analytis}},
  \bibinfo{author}{\bibfnamefont{R.~D.} \bibnamefont{McDonald}},
  \bibinfo{author}{\bibfnamefont{S.~C.} \bibnamefont{Riggs}},
  \bibinfo{author}{\bibfnamefont{J.-H.} \bibnamefont{Chu}},
  \bibinfo{author}{\bibfnamefont{G.~S.} \bibnamefont{Boebinger}},
  \bibnamefont{and} \bibinfo{author}{\bibfnamefont{I.~R.}
  \bibnamefont{Fisher}}, \bibinfo{journal}{Nature Physics}
  \textbf{\bibinfo{volume}{6}}, \bibinfo{pages}{960} (\bibinfo{year}{2010}).

\bibitem[{\citenamefont{Steinberg et~al.}(2010)\citenamefont{Steinberg,
  Gardner, Lee, and Jarillo-Herrero}}]{SteinbergHerreroNNano2010}
\bibinfo{author}{\bibfnamefont{H.}~\bibnamefont{Steinberg}},
  \bibinfo{author}{\bibfnamefont{D.~R.} \bibnamefont{Gardner}},
  \bibinfo{author}{\bibfnamefont{Y.~S.} \bibnamefont{Lee}}, \bibnamefont{and}
  \bibinfo{author}{\bibfnamefont{P.}~\bibnamefont{Jarillo-Herrero}},
  \bibinfo{journal}{Nano Letters} \textbf{\bibinfo{volume}{10}},
  \bibinfo{pages}{5032} (\bibinfo{year}{2010}).

\bibitem[{\citenamefont{Kim et~al.}(2012)\citenamefont{Kim, Cho, Butch, Syers,
  Kirshenbaum, Adam, Paglione, and Fuhrer}}]{Dohun_Fuhrer2012}
\bibinfo{author}{\bibfnamefont{D.}~\bibnamefont{Kim}},
  \bibinfo{author}{\bibfnamefont{S.}~\bibnamefont{Cho}},
  \bibinfo{author}{\bibfnamefont{N.~P.} \bibnamefont{Butch}},
  \bibinfo{author}{\bibfnamefont{P.}~\bibnamefont{Syers}},
  \bibinfo{author}{\bibfnamefont{K.}~\bibnamefont{Kirshenbaum}},
  \bibinfo{author}{\bibfnamefont{S.}~\bibnamefont{Adam}},
  \bibinfo{author}{\bibfnamefont{J.}~\bibnamefont{Paglione}}, \bibnamefont{and}
  \bibinfo{author}{\bibfnamefont{M.~S.} \bibnamefont{Fuhrer}},
  \bibinfo{journal}{Nature Physics} \textbf{\bibinfo{volume}{8}},
  \bibinfo{pages}{460} (\bibinfo{year}{2012}).

\bibitem[{\citenamefont{Bansal et~al.}(2011{\natexlab{a}})\citenamefont{Bansal,
  Kim, Brahlek, Edrey, and Oh}}]{Oh_arxiv2011}
\bibinfo{author}{\bibfnamefont{N.}~\bibnamefont{Bansal}},
  \bibinfo{author}{\bibfnamefont{Y.~S.} \bibnamefont{Kim}},
  \bibinfo{author}{\bibfnamefont{M.}~\bibnamefont{Brahlek}},
  \bibinfo{author}{\bibfnamefont{E.}~\bibnamefont{Edrey}}, \bibnamefont{and}
  \bibinfo{author}{\bibfnamefont{S.}~\bibnamefont{Oh}},
  \bibinfo{journal}{arXiv:1104.5709}  (\bibinfo{year}{2011}{\natexlab{a}}).

\bibitem[{\citenamefont{Sushkov et~al.}(2010)\citenamefont{Sushkov, Jenkins,
  Schmadel, Butch, Paglione, and Drew}}]{Sushkov_PRB2010}
\bibinfo{author}{\bibfnamefont{A.~B.} \bibnamefont{Sushkov}},
  \bibinfo{author}{\bibfnamefont{G.~S.} \bibnamefont{Jenkins}},
  \bibinfo{author}{\bibfnamefont{D.~C.} \bibnamefont{Schmadel}},
  \bibinfo{author}{\bibfnamefont{N.~P.} \bibnamefont{Butch}},
  \bibinfo{author}{\bibfnamefont{J.}~\bibnamefont{Paglione}}, \bibnamefont{and}
  \bibinfo{author}{\bibfnamefont{H.~D.} \bibnamefont{Drew}},
  \bibinfo{journal}{Physical Review B} \textbf{\bibinfo{volume}{82}},
  \bibinfo{pages}{125110} (\bibinfo{year}{2010}).

\bibitem[{\citenamefont{Brune et~al.}(2011)\citenamefont{Brune, Liu, Novik,
  Hankiewicz, Buhmann, Chen, Qi, Shen, Zhang, and Molenkamp}}]{Molenkamp2011}
\bibinfo{author}{\bibfnamefont{C.}~\bibnamefont{Brune}},
  \bibinfo{author}{\bibfnamefont{C.~X.} \bibnamefont{Liu}},
  \bibinfo{author}{\bibfnamefont{E.~G.} \bibnamefont{Novik}},
  \bibinfo{author}{\bibfnamefont{E.~M.} \bibnamefont{Hankiewicz}},
  \bibinfo{author}{\bibfnamefont{H.}~\bibnamefont{Buhmann}},
  \bibinfo{author}{\bibfnamefont{Y.~L.} \bibnamefont{Chen}},
  \bibinfo{author}{\bibfnamefont{X.~L.} \bibnamefont{Qi}},
  \bibinfo{author}{\bibfnamefont{Z.~X.} \bibnamefont{Shen}},
  \bibinfo{author}{\bibfnamefont{S.~C.} \bibnamefont{Zhang}}, \bibnamefont{and}
  \bibinfo{author}{\bibfnamefont{L.~W.} \bibnamefont{Molenkamp}},
  \bibinfo{journal}{Physical Review Letters} \textbf{\bibinfo{volume}{106}},
  \bibinfo{pages}{126803} (\bibinfo{year}{2011}).

\bibitem[{\citenamefont{Hancock et~al.}(2011)\citenamefont{Hancock, van
  Mechelen, Kuzmenko, van~der Marel, Brüne, Novik, Astakhov, Buhmann, and
  Molenkamp}}]{HancockMolenkamp2011}
\bibinfo{author}{\bibfnamefont{J.~N.} \bibnamefont{Hancock}},
  \bibinfo{author}{\bibfnamefont{J.~L.~M.} \bibnamefont{van Mechelen}},
  \bibinfo{author}{\bibfnamefont{A.~B.} \bibnamefont{Kuzmenko}},
  \bibinfo{author}{\bibfnamefont{D.}~\bibnamefont{van~der Marel}},
  \bibinfo{author}{\bibfnamefont{C.}~\bibnamefont{Brüne}},
  \bibinfo{author}{\bibfnamefont{E.~G.} \bibnamefont{Novik}},
  \bibinfo{author}{\bibfnamefont{G.~V.} \bibnamefont{Astakhov}},
  \bibinfo{author}{\bibfnamefont{H.}~\bibnamefont{Buhmann}}, \bibnamefont{and}
  \bibinfo{author}{\bibfnamefont{L.}~\bibnamefont{Molenkamp}},
  \bibinfo{journal}{1105.0884}  (\bibinfo{year}{2011}),
  \urlprefix\url{http://arxiv.org/abs/1105.0884}.

\bibitem[{\citenamefont{Kvon et~al.}(2012)\citenamefont{Kvon, Danilov, Kozlov,
  Zoth, Mikhailov, Dvoretskii, and Ganichev}}]{KvonGanichev2012}
\bibinfo{author}{\bibfnamefont{Z.}~\bibnamefont{Kvon}},
  \bibinfo{author}{\bibfnamefont{S.}~\bibnamefont{Danilov}},
  \bibinfo{author}{\bibfnamefont{D.}~\bibnamefont{Kozlov}},
  \bibinfo{author}{\bibfnamefont{C.}~\bibnamefont{Zoth}},
  \bibinfo{author}{\bibfnamefont{N.}~\bibnamefont{Mikhailov}},
  \bibinfo{author}{\bibfnamefont{S.}~\bibnamefont{Dvoretskii}},
  \bibnamefont{and} \bibinfo{author}{\bibfnamefont{S.}~\bibnamefont{Ganichev}},
  \bibinfo{journal}{JETP Letters} \textbf{\bibinfo{volume}{94}},
  \bibinfo{pages}{816–819} (\bibinfo{year}{2012}).

\bibitem[{\citenamefont{Schafgans et~al.}(2012)\citenamefont{Schafgans, Post,
  Taskin, Ando, Qi, Chapler, and Basov}}]{SchafgansBasov2012}
\bibinfo{author}{\bibfnamefont{A.~A.} \bibnamefont{Schafgans}},
  \bibinfo{author}{\bibfnamefont{K.~W.} \bibnamefont{Post}},
  \bibinfo{author}{\bibfnamefont{A.~A.} \bibnamefont{Taskin}},
  \bibinfo{author}{\bibfnamefont{Y.}~\bibnamefont{Ando}},
  \bibinfo{author}{\bibfnamefont{X.-L.} \bibnamefont{Qi}},
  \bibinfo{author}{\bibfnamefont{B.~C.} \bibnamefont{Chapler}},
  \bibnamefont{and} \bibinfo{author}{\bibfnamefont{D.~N.} \bibnamefont{Basov}},
  \bibinfo{journal}{Physical Review B} \textbf{\bibinfo{volume}{85}},
  \bibinfo{pages}{195440} (\bibinfo{year}{2012}).

\bibitem[{\citenamefont{Valdes~Aguilar
  et~al.}(2012)\citenamefont{Valdes~Aguilar, Stier, Liu, Bilbro, George,
  Bansal, Wu, Cerne, Markelz, Oh et~al.}}]{AguilarPRL2012}
\bibinfo{author}{\bibfnamefont{R.}~\bibnamefont{Valdes~Aguilar}},
  \bibinfo{author}{\bibfnamefont{A.~V.} \bibnamefont{Stier}},
  \bibinfo{author}{\bibfnamefont{W.}~\bibnamefont{Liu}},
  \bibinfo{author}{\bibfnamefont{L.~S.} \bibnamefont{Bilbro}},
  \bibinfo{author}{\bibfnamefont{D.~K.} \bibnamefont{George}},
  \bibinfo{author}{\bibfnamefont{N.}~\bibnamefont{Bansal}},
  \bibinfo{author}{\bibfnamefont{L.}~\bibnamefont{Wu}},
  \bibinfo{author}{\bibfnamefont{J.}~\bibnamefont{Cerne}},
  \bibinfo{author}{\bibfnamefont{A.~G.} \bibnamefont{Markelz}},
  \bibinfo{author}{\bibfnamefont{S.}~\bibnamefont{Oh}}, \bibnamefont{et~al.},
  \bibinfo{journal}{Physical Review Letters} \textbf{\bibinfo{volume}{108}},
  \bibinfo{pages}{087403} (\bibinfo{year}{2012}).

\bibitem[{\citenamefont{Bansal et~al.}(2011{\natexlab{b}})\citenamefont{Bansal,
  Kim, Edrey, Brahlek, Horibe, Iida, Tanimura, Li, Feng, Lee
  et~al.}}]{Bansal_OhThinFilm2011}
\bibinfo{author}{\bibfnamefont{N.}~\bibnamefont{Bansal}},
  \bibinfo{author}{\bibfnamefont{Y.~S.} \bibnamefont{Kim}},
  \bibinfo{author}{\bibfnamefont{E.}~\bibnamefont{Edrey}},
  \bibinfo{author}{\bibfnamefont{M.}~\bibnamefont{Brahlek}},
  \bibinfo{author}{\bibfnamefont{Y.}~\bibnamefont{Horibe}},
  \bibinfo{author}{\bibfnamefont{K.}~\bibnamefont{Iida}},
  \bibinfo{author}{\bibfnamefont{M.}~\bibnamefont{Tanimura}},
  \bibinfo{author}{\bibfnamefont{G.-H.} \bibnamefont{Li}},
  \bibinfo{author}{\bibfnamefont{T.}~\bibnamefont{Feng}},
  \bibinfo{author}{\bibfnamefont{H.-D.} \bibnamefont{Lee}},
  \bibnamefont{et~al.}, \bibinfo{journal}{Thin Solid Films}
  \textbf{\bibinfo{volume}{520}}, \bibinfo{pages}{224}
  (\bibinfo{year}{2011}{\natexlab{b}}).

\bibitem[{\citenamefont{Jenkins
  et~al.}(2010{\natexlab{b}})\citenamefont{Jenkins, Schmadel, and
  Drew}}]{Jenkins_RSI_2010}
\bibinfo{author}{\bibfnamefont{G.~S.} \bibnamefont{Jenkins}},
  \bibinfo{author}{\bibfnamefont{D.~C.} \bibnamefont{Schmadel}},
  \bibnamefont{and} \bibinfo{author}{\bibfnamefont{H.~D.} \bibnamefont{Drew}},
  \bibinfo{journal}{Review of Scientific Instruments}
  \textbf{\bibinfo{volume}{81}}, \bibinfo{pages}{083903}
  (\bibinfo{year}{2010}{\natexlab{b}}).

\bibitem[{\citenamefont{O'Connell and Wallace}(1982)}]{OConnellWallace1982}
\bibinfo{author}{\bibfnamefont{R.~F.} \bibnamefont{O'Connell}}
  \bibnamefont{and} \bibinfo{author}{\bibfnamefont{G.}~\bibnamefont{Wallace}},
  \bibinfo{journal}{Physical Review B} \textbf{\bibinfo{volume}{26}},
  \bibinfo{pages}{2231} (\bibinfo{year}{1982}).

\bibitem[{\citenamefont{Morimoto et~al.}(2009)\citenamefont{Morimoto, Hatsugai,
  and Aoki}}]{MorimotoAoki2009}
\bibinfo{author}{\bibfnamefont{T.}~\bibnamefont{Morimoto}},
  \bibinfo{author}{\bibfnamefont{Y.}~\bibnamefont{Hatsugai}}, \bibnamefont{and}
  \bibinfo{author}{\bibfnamefont{H.}~\bibnamefont{Aoki}},
  \bibinfo{journal}{Physical Review Letters} \textbf{\bibinfo{volume}{103}},
  \bibinfo{pages}{116803} (\bibinfo{year}{2009}).

\bibitem[{\citenamefont{Stier et~al.}(2011)\citenamefont{Stier, Zhang, Ellis,
  Eason, Strasser, Morimoto, McCombe, Aoki, and Cerne}}]{StierCerne2011}
\bibinfo{author}{\bibfnamefont{A.~V.} \bibnamefont{Stier}},
  \bibinfo{author}{\bibfnamefont{H.}~\bibnamefont{Zhang}},
  \bibinfo{author}{\bibfnamefont{C.~T.} \bibnamefont{Ellis}},
  \bibinfo{author}{\bibfnamefont{D.}~\bibnamefont{Eason}},
  \bibinfo{author}{\bibfnamefont{G.}~\bibnamefont{Strasser}},
  \bibinfo{author}{\bibfnamefont{T.}~\bibnamefont{Morimoto}},
  \bibinfo{author}{\bibfnamefont{B.~D.} \bibnamefont{McCombe}},
  \bibinfo{author}{\bibfnamefont{H.}~\bibnamefont{Aoki}}, \bibnamefont{and}
  \bibinfo{author}{\bibfnamefont{J.}~\bibnamefont{Cerne}},
  \bibinfo{journal}{arXiv:1201.0182}  (\bibinfo{year}{2011}),
  \urlprefix\url{http://arxiv.org/abs/1201.0182}.

\bibitem[{\citenamefont{Xia et~al.}(2009)\citenamefont{Xia, Qian, Hsieh, Wray,
  Pal, Lin, Bansil, Grauer, Hor, Cava et~al.}}]{XiaARPESNP2009}
\bibinfo{author}{\bibfnamefont{Y.}~\bibnamefont{Xia}},
  \bibinfo{author}{\bibfnamefont{D.}~\bibnamefont{Qian}},
  \bibinfo{author}{\bibfnamefont{D.}~\bibnamefont{Hsieh}},
  \bibinfo{author}{\bibfnamefont{L.}~\bibnamefont{Wray}},
  \bibinfo{author}{\bibfnamefont{A.}~\bibnamefont{Pal}},
  \bibinfo{author}{\bibfnamefont{H.}~\bibnamefont{Lin}},
  \bibinfo{author}{\bibfnamefont{A.}~\bibnamefont{Bansil}},
  \bibinfo{author}{\bibfnamefont{D.}~\bibnamefont{Grauer}},
  \bibinfo{author}{\bibfnamefont{Y.~S.} \bibnamefont{Hor}},
  \bibinfo{author}{\bibfnamefont{R.~J.} \bibnamefont{Cava}},
  \bibnamefont{et~al.}, \bibinfo{journal}{Nature Physics}
  \textbf{\bibinfo{volume}{5}}, \bibinfo{pages}{398} (\bibinfo{year}{2009}).

\bibitem[{\citenamefont{Zhu et~al.}(2011)\citenamefont{Zhu, Levy, Ludbrook,
  Veenstra, Rosen, Comin, Wong, Dosanjh, Ubaldini, Syers et~al.}}]{ZhuPRL2011}
\bibinfo{author}{\bibfnamefont{Z.-H.} \bibnamefont{Zhu}},
  \bibinfo{author}{\bibfnamefont{G.}~\bibnamefont{Levy}},
  \bibinfo{author}{\bibfnamefont{B.}~\bibnamefont{Ludbrook}},
  \bibinfo{author}{\bibfnamefont{C.~N.} \bibnamefont{Veenstra}},
  \bibinfo{author}{\bibfnamefont{J.~A.} \bibnamefont{Rosen}},
  \bibinfo{author}{\bibfnamefont{R.}~\bibnamefont{Comin}},
  \bibinfo{author}{\bibfnamefont{D.}~\bibnamefont{Wong}},
  \bibinfo{author}{\bibfnamefont{P.}~\bibnamefont{Dosanjh}},
  \bibinfo{author}{\bibfnamefont{A.}~\bibnamefont{Ubaldini}},
  \bibinfo{author}{\bibfnamefont{P.}~\bibnamefont{Syers}},
  \bibnamefont{et~al.}, \bibinfo{journal}{Physical Review Letters}
  \textbf{\bibinfo{volume}{107}}, \bibinfo{pages}{186405}
  (\bibinfo{year}{2011}).

\bibitem[{\citenamefont{Bahramy et~al.}(2012)\citenamefont{Bahramy, King, de~la
  Torre, Chang, Shi, Patthey, Balakrishnan, Hofmann, Arita, Nagaosa
  et~al.}}]{Bahramy_Hofmann_2012}
\bibinfo{author}{\bibfnamefont{M.~S.} \bibnamefont{Bahramy}},
  \bibinfo{author}{\bibfnamefont{P.~D.~C.} \bibnamefont{King}},
  \bibinfo{author}{\bibfnamefont{A.}~\bibnamefont{de~la Torre}},
  \bibinfo{author}{\bibfnamefont{J.}~\bibnamefont{Chang}},
  \bibinfo{author}{\bibfnamefont{M.}~\bibnamefont{Shi}},
  \bibinfo{author}{\bibfnamefont{L.}~\bibnamefont{Patthey}},
  \bibinfo{author}{\bibfnamefont{G.}~\bibnamefont{Balakrishnan}},
  \bibinfo{author}{\bibfnamefont{P.}~\bibnamefont{Hofmann}},
  \bibinfo{author}{\bibfnamefont{R.}~\bibnamefont{Arita}},
  \bibinfo{author}{\bibfnamefont{N.}~\bibnamefont{Nagaosa}},
  \bibnamefont{et~al.}, \bibinfo{journal}{arXiv:1206.0564}
  (\bibinfo{year}{2012}), \urlprefix\url{http://arxiv.org/abs/1206.0564}.

\bibitem[{\citenamefont{Cao et~al.}(2012)\citenamefont{Cao, Tian, Miotkowski,
  Shen, Hu, Qiao, and Chen}}]{CaoChen2012}
\bibinfo{author}{\bibfnamefont{H.}~\bibnamefont{Cao}},
  \bibinfo{author}{\bibfnamefont{J.}~\bibnamefont{Tian}},
  \bibinfo{author}{\bibfnamefont{I.}~\bibnamefont{Miotkowski}},
  \bibinfo{author}{\bibfnamefont{T.}~\bibnamefont{Shen}},
  \bibinfo{author}{\bibfnamefont{J.}~\bibnamefont{Hu}},
  \bibinfo{author}{\bibfnamefont{S.}~\bibnamefont{Qiao}}, \bibnamefont{and}
  \bibinfo{author}{\bibfnamefont{Y.~P.} \bibnamefont{Chen}},
  \bibinfo{journal}{Physical Review Letters} \textbf{\bibinfo{volume}{108}},
  \bibinfo{pages}{216803} (\bibinfo{year}{2012}).

\bibitem[{\citenamefont{Haavasoja et~al.}(1984)\citenamefont{Haavasoja,
  Stormer, Bishop, Narayanamurti, Gossard, and Wiegmann}}]{Haavasoja1984}
\bibinfo{author}{\bibfnamefont{T.}~\bibnamefont{Haavasoja}},
  \bibinfo{author}{\bibfnamefont{H.}~\bibnamefont{Stormer}},
  \bibinfo{author}{\bibfnamefont{D.}~\bibnamefont{Bishop}},
  \bibinfo{author}{\bibfnamefont{V.}~\bibnamefont{Narayanamurti}},
  \bibinfo{author}{\bibfnamefont{A.}~\bibnamefont{Gossard}}, \bibnamefont{and}
  \bibinfo{author}{\bibfnamefont{W.}~\bibnamefont{Wiegmann}},
  \bibinfo{journal}{Surface Science} \textbf{\bibinfo{volume}{142}},
  \bibinfo{pages}{294–297} (\bibinfo{year}{1984}).

\bibitem[{\citenamefont{Kohler}(1973)}]{Köhler1973}
\bibinfo{author}{\bibfnamefont{H.}~\bibnamefont{Kohler}},
  \bibinfo{journal}{physica status solidi (b)} \textbf{\bibinfo{volume}{58}},
  \bibinfo{pages}{91–100} (\bibinfo{year}{1973}).

\bibitem[{\citenamefont{Eto et~al.}(2010)\citenamefont{Eto, Ren, Taskin,
  Segawa, and Ando}}]{EtoAndo2010}
\bibinfo{author}{\bibfnamefont{K.}~\bibnamefont{Eto}},
  \bibinfo{author}{\bibfnamefont{Z.}~\bibnamefont{Ren}},
  \bibinfo{author}{\bibfnamefont{A.~A.} \bibnamefont{Taskin}},
  \bibinfo{author}{\bibfnamefont{K.}~\bibnamefont{Segawa}}, \bibnamefont{and}
  \bibinfo{author}{\bibfnamefont{Y.}~\bibnamefont{Ando}},
  \bibinfo{journal}{Physical Review B} \textbf{\bibinfo{volume}{81}},
  \bibinfo{pages}{195309} (\bibinfo{year}{2010}).

\end{thebibliography}

\end{document}